\documentclass[twocolumn,prl,amsmath,amssymb,showpacs]{revtex4}
\usepackage{graphicx}
\usepackage{dcolumn}
\usepackage{bm}
\usepackage{latexsym}
\usepackage{amstext}
\usepackage{amssymb}
\usepackage{amsxtra}
\usepackage[english]{babel}

\begin{document}
\title{Critical Point of an Interacting Two-Dimensional Atomic Bose Gas}

\author{Peter Kr\"uger}
\author{Zoran Hadzibabic}
\author{Jean Dalibard}
\affiliation{Laboratoire Kastler Brossel and CNRS, Ecole Normale
Sup\'erieure, 24 rue Lhomond, 75005 Paris, France}
\date{\today}

\begin{abstract}

We have measured the critical atom number in an array of
harmonically trapped two-dimensional (2D) Bose gases of rubidium
atoms at different temperatures. We found this number to be about
five times higher than predicted by the semi-classical theory of
Bose-Einstein condensation (BEC) in the ideal gas. This demonstrates
that the conventional BEC picture is inapplicable in an interacting
2D atomic gas, in sharp contrast to the three-dimensional case. A
simple heuristic model based on the Berezinskii-Kosterlitz-Thouless
theory of 2D superfluidity and the local density approximation
accounts well for our experimental results.

\end{abstract}


\maketitle

Bose-Einstein condensation (BEC) at a finite temperature is not
possible in a homogeneous two-dimensional (2D) system, but an
interacting Bose fluid can nevertheless become superfluid at a
finite critical temperature~\cite{Bishop:1978}. This unconventional phase
transition is described by the Berezinskii-Kosterlitz-Thouless (BKT)
theory~\cite{Berezinskii:1971,Kosterlitz:1973}, and does not involve any spontaneous symmetry
breaking and emergence of a uniform order parameter. It is instead
associated with a topological order embodied in the pairing of
vortices with opposite circulations; true long-range order is
destroyed by long wavelength phase fluctuations even in the
superfluid state~\cite{Mermin:1966,Hohenberg:1967}.

Recent advances in producing harmonically trapped, weakly
interacting (quasi-)2D atomic gases~\cite{Gorlitz:2001, Burger:2002,
rych04, Hadzibabic:2004, smit05, Kohl:2005b, Stock:2005,
Hadzibabic:2006, spiel07} have opened the possibility for detailed
studies of BKT physics in a controllable environment. There has been
some theoretical debate on the nature of the superfluid transition
in these systems~\cite{petr00BEC2D, Petrov:2001, ande02,
Simula:2006, Holzmann:2005} because the harmonic confinement
modifies the density of states compared to the homogenous case. This
allows for ``conventional" finite temperature Bose-Einstein
condensation in the \textit{ideal} 2D gas~\cite{Bagnato:1991}. Early
experiments have been equally consistent with the BEC and the BKT
picture of the phase transition. For example, the density profiles
at very low temperatures~\cite{Gorlitz:2001} are expected to be the
same in both cases. However, recent studies of matter wave
interference of independent 2D atomic clouds close to the transition
have revealed both thermally activated vortices~\cite{Stock:2005,
Hadzibabic:2006} and quasi-long-range coherence
properties~\cite{Hadzibabic:2006} in agreement with the BKT
theory~\cite{Nelson:1977, Polkovnikov:2006}.

In this Letter, we study the critical atom number in an array of
2D gases of rubidium atoms, and observe stark disagreement
with the predictions of the ideal gas BEC theory. We detect the
critical point by measuring (i) the onset of
bimodality in the atomic density distribution and (ii) the onset of
interference between independent 2D clouds. These two measurements
agree with each other, and for the investigated range of temperatures $T \approx
50$--110\,nK give critical atom numbers $N_{\rm c}$ which are $\sim
5$ times higher than the ideal gas prediction for conventional
Bose-Einstein condensation in our trap~\cite{Bagnato:1991}. For
comparison, in three-dimensional (3D) atomic gases, where
conventional BEC occurs, the increase of the critical atom number
due to repulsive interactions is typically on the order of ten
percent \cite{Dalfovo:1999, Gerbier:2004}. A simple
heuristic model based on the BKT theory of 2D superfluidity and the
local density approximation gives good agreement with our
measurements.

\begin{figure}
\vspace{-2mm} \centerline{\includegraphics[angle=-90,
width=\columnwidth]{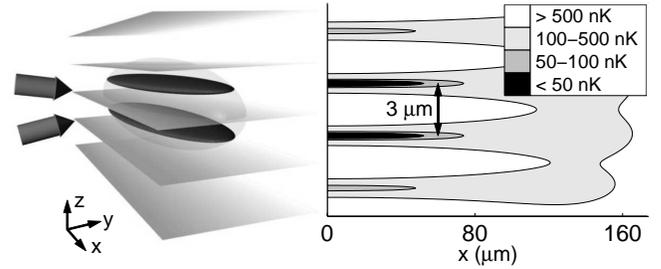}} \vspace{-2mm} \caption{
Experimental setup. Left: A one-dimensional optical lattice is used
to split a magnetically trapped 3D BEC (transparent ellipsoid) into
a small array of 2D clouds. Right: Contour lines of the total (magnetic and light) potential $V(x,z)$
in the $y=0$ plane for the lattice phase such that the two
central planes are symmetric with respect to the trap center.}
 \label{fig:contour}\vspace{-5mm}
\end{figure}
\begin{figure}
\vspace{-7mm} \centerline{\includegraphics[angle=-90,
width=\columnwidth]{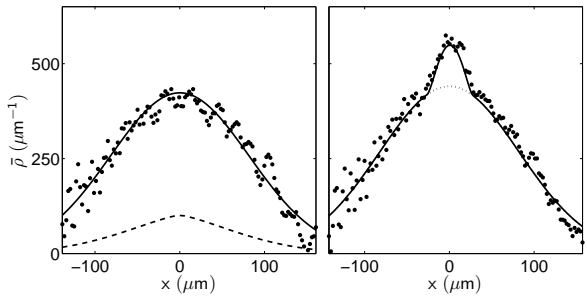}} \vspace{-7mm} \caption{Phase
transition in a rubidium 2D gas. 2D clouds confined parallel to the
$xy$ plane are released from an optical lattice and the density
distribution is recorded by absorption imaging along $y$ after
$t=22\,$ms of time of flight. The measured line densities
$\bar{\rho}(x)$ ($\bullet$) for an atom number just below (left) and
just above (right) the critical number are displayed together with
bimodal fits (solid lines). The dashed line in the left panel shows
the expected distribution of the 2D ideal gas at the threshold of
conventional BEC in our potential at the same temperature
($T = 92\,$nK). The dotted line in the right panel indicates the
Gaussian part of the bimodal distribution.} \label{fig:bimodality}
\vspace{-5mm}
\end{figure}

In~\cite{Hadzibabic:2006} we studied quasi-long-range coherence of a
trapped 2D gas, which is directly related to the \textit{superfluid}
density $\rho_{\rm s}$~\cite{Polkovnikov:2006}. In that case,
signatures of the BKT transition emerge only once a significant part
of the cloud becomes superfluid. Since the atomic density in the
trap is not uniform, this happens slightly below the true critical
temperature for the onset of superfluidity in the trap center, and
the observed transition is rounded off. The present study
concentrates on the exact critical point and relates to the
\textit{total} density at criticality $\rho_{\rm c}$, which has been
of long standing theoretical
interest~\cite{Fisher:1988,Prokof'ev:2001}.

Our experimental procedure for the preparation of cold 2D Bose gases
has been described in \cite{Hadzibabic:2006}. We start with a
$^{87}$Rb 3D condensate in a cylindrically symmetric magnetic trap
with trapping frequencies $\omega_x=2\pi\times 10.6\,$Hz and
$\omega_y=\omega_z=2\pi\times 125\,$Hz. To split the sample into 2D
clouds we add a blue detuned one-dimensional optical lattice with a
period of $d = 3\,\mu$m along the vertical direction $z$ (see Fig.~\ref{fig:contour}).
The lattice is formed by two laser beams with a $532\,$nm wavelength and
focussed to waists of about $120\,\mu$m, which propagate in the
$yz$ plane and intersect at a small angle. The depth of the lattice
potential around $x=0$ is $h\times 35\,$kHz, corresponding to a
vertical confinement of $\omega_z=2\pi\times 3.0\,$kHz. The
tunneling rate between adjacent sites at the center of the trap
($x=0$) is negligible on the time scale of the experiment. The
finite waists of the lattice beams result in a slow variation of
$\omega_z$ along $x$, and the variation of the zero point energy
$\hbar\omega_z(x)/2$ modifies $\omega_x$ to $2\pi \times 9.4\,$Hz at
the trap center.

Fig.~\ref{fig:contour} shows contour lines for the full trapping
potential. The number of significantly populated lattice planes is
$\sim 2 - 4$ in the investigated temperature range (50-110~nK). The vast majority
of atoms is trapped in the central $x$ region where the 2D criterion
$kT<\hbar \omega_z(x)$ is fulfilled. However, the exchange of
particles between lattice sites is still possible via the far wings
of the energy distribution (at energies above 460~nK). This ensures
thermal equilibrium between the planes~\cite{shin04dis} on the time
scale of $\sim 100$ collision times \cite{Luiten:1996}, which in our
case corresponds to a fraction of a second. The 2D interaction
strength is $g=(\hbar^2/m)\tilde{g}$, where the dimensionless
coupling constant $\tilde{g}=a_{\rm s}\sqrt{8\pi m \omega_z/\hbar} =
0.13$, $a_s = 5.2\,$nm is the scattering length, and $m$ the atomic
mass \cite{petr00BEC2D,Kagan:1987}. The interaction energy $E_{\rm
int} \sim g \rho_0$, where $\rho_0$ is the peak density,
also satisfies the 2D criterion $E_{\rm int}<\hbar\omega_z$.

We measure the critical atom number $N_{\rm c}$ by varying the total
atom number $N$ at a fixed temperature. We start with a highly
degenerate sample and keep it trapped for a time $\tau$ varying
between 1 and 10\,s. During this time we maintain a constant
temperature by applying a constant radio frequency field in the
range of 10--25\,kHz above the frequency corresponding to the bottom
of the trap. As the hold time $\tau$
increases, $N$ gradually  reduces and drops below $N_{\rm c}$ due to
inelastic losses.

The atomic density profiles are recorded in the $xz$ plane by
resonant absorption imaging along $y$ after $t=22\,$ms of time of
flight expansion.  Along $z$ the profiles are Gaussian, closely
corresponding to the zero point kinetic energy
$\hbar\omega_z(x=0)/4$. Along $x$, for all $N<N_{\rm c}$ a Gaussian
distribution fits the data well~\cite{footnote_y}. For $N>N_{\rm
c}$, the profiles exhibit a clearly bimodal shape
(Fig.~\ref{fig:bimodality}). The bimodal distributions are fitted
well by the sum of a Gaussian, corresponding to the ``normal
component", and a parabolic Thomas-Fermi (TF) profile expected from
superfluid hydrodynamics~\cite{Dalfovo:1999}.

\begin{figure}
\vspace{1mm} \centerline{\includegraphics[angle=-90,
width=\columnwidth]{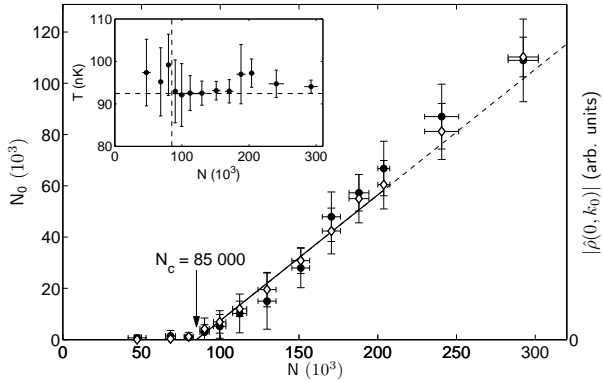}} \vspace{-2mm} \caption{Measurement
of the critical point. The number of atoms in the Thomas-Fermi part
of the bimodal distribution $N_0$ ($\diamond$) is plotted as a
function of the total atom number $N$. The solid line shows the
linear fit we use to determine $N_{\rm c}$, and the dashed line is
its extrapolation. For comparison, the interference amplitude
$|\hat{\rho}(0, k_0)|$ ($\bullet$) is also displayed as a function
of $N$. It shows the same threshold $N_{\rm c}$ within our
experimental precision. The inset shows that the temperature deduced
from the Gaussian part of the fit is to a good approximation
constant for all data points. Horizontal and vertical dashed lines
indicate the average temperature and the critical atom number,
respectively. The solid line marks the region used to determine the
average temperature $T=92(6)\,$nK close to the transition. Each data
point is based on 5--10 images, all error bars represent standard
deviations.} \label{fig:laser} \vspace{-5mm}
\end{figure}

From the bimodal fits we extract the total atom number $N$ and the
number of atoms within the TF part of the distribution $N_0$. The
absolute detection efficiency of our imaging system was calibrated
by measuring critical atom numbers for 3D BECs, taking into account
interaction effects \cite{Dalfovo:1999, Gerbier:2004}. For a given
energy of the evaporation surface $E_\mathrm{evap}$ the width of the
Gaussian part of the distribution is nearly independent of $N$ (see
inset of Fig.~\ref{fig:laser}). For a quasi-non degenerate gas
($N\sim N_{\rm c}/2$) this width is given by the temperature. We
thus use this estimate for $T$ also in the degenerate regime ($N
\gtrsim N_{\rm c}$), although one could have expected in this regime
a deviation from the Gaussian law for the normal fraction. For
example the ideal gas theory predicts a distribution at the BEC
point that is much more peaked at the center of the cloud
(Fig.~\ref{fig:bimodality}). The temperatures inferred in this way
scale as $E_\mathrm{evap} = \eta k T$, with $\eta\approx 10$
compatible with the usual 3D values for evaporation times equal to a
few hundred collision times~\cite{Luiten:1996}. We estimate the
systematic uncertainties of our atom number and temperature
calibrations to be 20\% and 10\%, respectively.

\begin{figure}
\vspace{1mm} \centerline{\includegraphics[angle=-90,
width=\columnwidth]{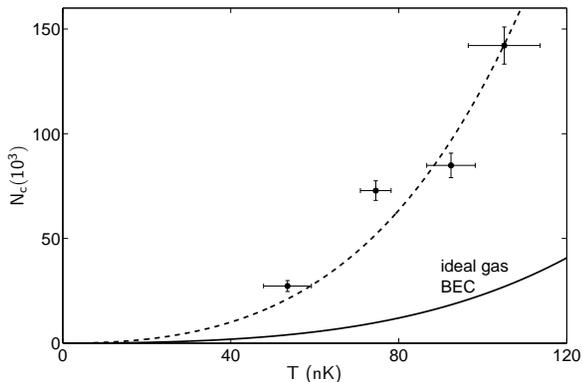}} \vspace{-4mm} \caption{Critical
point in an interacting 2D gas. The critical atom number $N_{\rm c}$
($\bullet$) is measured at four different temperatures $T$.
Displayed error bars are statistical. The solid line shows the ideal
2D gas BEC prediction $N^{\rm multi}_{\rm c,id}$. The dashed line is
the best empirical fit to the data, which gives $N_{\rm c} =
\alpha\,N_{\rm c,id}^{\rm multi}$ with $\alpha=5.3(5)$.}
\label{fig:NcvsT} \vspace{-5mm}
\end{figure}

Fig.~\ref{fig:laser} illustrates the threshold behavior of
$N_0$~\cite{footnote_slope}, and Fig.~\ref{fig:NcvsT} shows the
critical numbers $N_{\rm c}$ measured at four different
temperatures. In a single 2D ideal gas, BEC is expected for
 \cite{Bagnato:1991}:
 \begin{equation}
N_{\rm c,id} =
\frac{\pi^2}{6}\left(\frac{kT}{\hbar\bar{\omega}}\right)^2 \ ,
 \label{eq:Ncideal}
 \end{equation}
where $\bar{\omega}$ is the geometric mean of the two trapping
frequencies in the plane. For comparison with our experimental
results we have numerically integrated the Bose-Einstein
distribution for our confining potential sketched in
Fig.~\ref{fig:contour}. The result depends on a ten percent level on
the exact position of the lattice planes relative to the minimum of
the magnetic trap potential. Since we do not fully control this
position, we average over the possible configurations. We obtain the
result $N_{\rm c,id}^{\rm multi} = p N_{\rm c,id}$, where the
effective number of planes $p$ smoothly grows from $\approx 2.2$ at
$50\,$nK to $\approx4.2$ at $110\,$nK. The resulting $N_{\rm
c,id}^{\rm multi}(T)$ is shown in Fig.~\ref{fig:NcvsT} as a solid
line. Our measurements clearly show systematically higher $N_{\rm
c}$ than expected for ideal gas condensation. An empirical function
$N_{\rm c} = \alpha N_{\rm c,id}^{\rm multi}(T)$, with the scaling
factor $\alpha$ as the only free parameter, fits the data well and
gives $\alpha =5.3(5)$, where the quoted error is statistical.

\begin{figure}
\vspace{-7mm} \centerline{\includegraphics[angle=-90, width=1.1
\columnwidth]{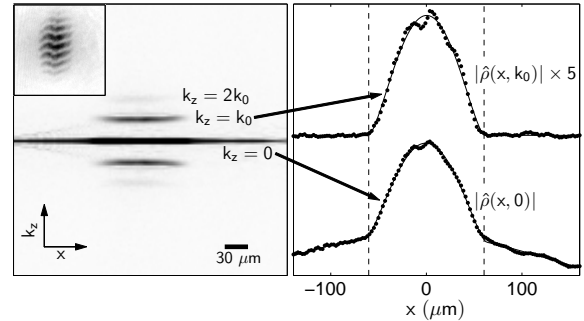}} \vspace{-7mm} \caption{Coherence and
density profile below the transition. Interference of 2D clouds is
used to compare the coherent part of the cloud with the part
following the Thomas-Fermi density distribution. Left: Interference
patterns in the $xz$ plane (see example in inset) are Fourier
transformed along the expansion axis $z$ and averaged over ten
images taken under identical conditions, to obtain $|\hat{\rho}(x,
k_z)|$. Right: Within experimental precision, fits to the total
density profile $|\hat{\rho}(x,0)|$ and the interference amplitude
profile $|\hat{\rho}(x,k_0)|$ give the same Thomas-Fermi diameter $2
R_{\rm TF}$, indicated by the dashed lines. The weak second harmonic
peak at $k_z = 2 k_0$ reveals small occupation of the outer lattice
planes.} \label{fig:interference} \vspace{-5mm}
\end{figure}

We also study the coherent fraction of the 2D gas and compare its
behavior with the bimodal density profiles. We investigate the
interference patterns that form after releasing the independent
planar gases from the trap (Fig.\ \ref{fig:interference})
\cite{Hadzibabic:2006}. Fourier transforming the density profile
$\rho(x,z)\rightarrow \mathcal{F}[\rho(x,z)]
\equiv\hat{\rho}(x,k_z)$ allows us to quantify the size of the
coherent, i.e.\ interfering part of the gas as a function of $N$.
The spatial frequency corresponding to the fringe period for the
interference of neighboring planes is $k_0=md/\hbar t$. We find that
$\hat{\rho}(x,k_0)$ is well fitted by a pure Thomas-Fermi profile.
Within our experimental accuracy, the radii $R_\mathrm{TF}(k_0)$ of
these profiles are equal to those obtained from a bimodal fit to the
density. In particular, the onsets of interference and bimodality
coincide (circles and diamonds in Fig.~\ref{fig:laser},
respectively).

We now turn to the interpretation of our measurements in the
framework of the BKT theory of 2D superfluidity. The theory predicts
a universal jump of the \textit{superfluid} density at the
transition, from $\rho_{\rm s} = 0$  to $\rho_{\rm s}\lambda^2 = 4$,
where $\lambda = h/\sqrt{2\pi mkT}$ is the thermal
wavelength~\cite{Nelson:1977} (for experiments see
\cite{Bishop:1978, Hadzibabic:2006}). However, the \textit{total}
density at the critical point $\rho_{\rm c}$ is not
universal because it depends on the microscopic interactions. For
weak interactions ($\tilde{g}<1$),
$\rho_{\rm c} \lambda^2=\ln\left(C/\tilde{g}\right)$~\cite{Fisher:1988},
with $C=380 \pm 3$ given by high-precision Monte
Carlo calculations~\cite{Prokof'ev:2001}. For our
value of $\tilde{g}=0.13$ (experimentally confirmed by measuring
$R_{\rm TF}$ as a function of $N_0$) this gives $\rho_{\rm
c}\lambda^2 = 8.0$.

In a harmonic trap, within the local density approximation, the
transition is expected to occur when the density in the center of
the cloud reaches the critical value $\rho_{\rm c}$. We can
heuristically relate the critical density and the corresponding
critical atom number $N_{\rm c,BKT}$ using the experimentally
observed Gaussian density profiles. For a single plane with a
quadratic confining potential $V(x,y)$, integrating $\rho(x,y)=
\rho_{\rm c} \exp(-V(x,y)/(kT))$ gives:
\begin{equation}
N_{\rm c,BKT} = \rho_{\rm c}\lambda^2
\left(\frac{kT}{\hbar\bar{\omega}}\right)^2 = \rho_{\rm c}\lambda^2
\frac{6}{\pi^2} N_{\rm c,id} \ .
 \label{eq:Nc_BKT}
 \end{equation}
For $\rho_{\rm c}\lambda^2=8.0$ this gives $N_{\rm c,BKT}/N_{\rm
c,id}= 4.9$. In a lattice configuration this ratio changes only
slightly. We set the peak density in the most populated plane to
$\rho_{\rm c}$, and sum the contributions of all planes $j$ using
the corresponding potentials $V_j$ (here we neglect the small non-harmonic effects due to finite laser
waists). The total population in the
lattice is then $N_{\rm c,BKT}^{\rm multi}= p'N_{\rm c,BKT}$, where
the effective number of planes $p'$ varies from 2.4 at 50~nK to 3.5
at 110~nK. We thus obtain $N_{\rm c,BKT}^{\rm multi}/N_{\rm
c,id}^{\rm multi}\simeq 4.7$, which is close to the experimental
ratio $\alpha = 5.3$.

One could try to reproduce our observations within the
self-consistent Hartree-Fock (HF) theory~\cite{Holzmann:2005} (see
also \cite{Bhaduri:2000,Fernandez:2002,gies04}), by replacing
$V(\textbf{r})$ with the effective mean field potential
$V(\textbf{r}) + 2 g\rho({\bf r})$ and again setting the peak
density to the BKT threshold $\rho_{\rm c}$. For very weak
interactions, $\log(1/\tilde{g}) \gg 1$, analytical HF calculation
gives critical numbers which are only slightly larger than $N_{\rm
c, id}$~\cite{Holzmann:2005}. This approach could in principle be
implemented numerically for our value of $\tilde g$ and our
lattice geometry. However, it has been
suggested~\cite{Prokof'ev:2002} that treating interactions at the
mean field level is insufficient for $\tilde{g} \sim
10^{-1}$, because the interactions are strong enough for the
critical region to be a significant fraction of the sample. In
future experiments with atomic gases $\tilde{g}$ could be varied
between 1 and $10^{-4}$ using Feshbach resonances, allowing for
detailed tests of the microscopic BKT theory and the possible
breakdown of the mean field approximation.

In conclusion, we have shown that the ideal gas theory of
Bose-Einstein condensation, which is extremely successful in 3D,
cannot be used to predict the critical point in interacting 2D
atomic gases, where interactions play a profound role even in the normal state.
A much better prediction of the critical point is provided by the BKT theory of 2D superfluidity.
We have also shown that, despite the absence of true long-range
order, the low temperature state displays
density profiles and local coherence largely analogous to 3D BECs.

\acknowledgments We thank B. Battelier, M. Cheneau, P. Rath, D.
Stamper-Kurn, N. Cooper, and M. Holzmann for useful discussions.
P.K. and Z.H. acknowledge support from the EU (contracts
MEIF-CT-2006-025047 and MIF1-CT-2005-007932). This work is supported
by R\'egion Ile de France (IFRAF), CNRS, the French Ministry of
Research, and ANR. Laboratoire Kastler Brossel is a research unit of
Ecole Normale Sup\'{e}rieure, Universit\'e Pierre and Marie Curie
and CNRS.


\begin{thebibliography}{33}
\bibitem{footnote_y}
{By imaging along $x$, we verified that the profiles along $y$
are also fitted well by a Gaussian.}
\bibitem{footnote_slope}
{Note that the slope $dN_0/dN \sim 0.7$ for $N>N_{\rm c}$ is less
than unity, contrary to what is expected for a saturated gas. This
lack of saturation will be discussed elsewhere.}
\end{thebibliography}
\end{document}